\documentclass[useAMS,usegraphicx]{mn2e}

\title[Tidal origin of the Magellanic Stream]{Constraining the orbital history of the Magellanic Clouds: A new bound scenario suggested by the tidal origin of the Magellanic Stream}
\author[J. Diaz \& K. Bekki]{Jonathan Diaz$^1$\thanks{Email: jonathan.diaz@icrar.org} and Kenji Bekki$^1$ \\ $^1$ International Centre for Radio Astronomy Research, M468, The University of Western Australia, 35 Stirling Highway, Crawley, Western Australia 6009, Australia}

\begin{document}

\maketitle

\begin{abstract}

Bound orbits have traditionally been assigned to the Large and Small Magellanic Clouds (LMC and SMC, respectively) in order to provide a formation scenario for the Magellanic Stream (MS) and its Leading Arm (LA), two prominent neutral hydrogen (HI) features connected to the LMC and SMC.  However, Hubble Space Telescope (HST) measurements of the proper motions of the LMC and SMC have challenged the plausibility of bound orbits, causing the origin of the MS to re-emerge as a contested issue.  { We present a new tidal model in which structures resembling the bifurcated MS and elongated LA are able to form in a bound orbit consistent with the HST proper motions.  The LMC and SMC have remained bound to each other only recently in our model despite being separately bound to the Milky Way for more than 5 Gyr.  We find that the MS and LA are able to form as a consequence of LMC-dominated tidal stripping during the recent dynamical coupling of the LMC and SMC.  Our orbital model depends on our assumption that the Milky Way has a constant circular velocity of $V_{\rm cir}=$ 250 km s$^{-1}$ up to 160 kpc, which implies a massive isothermal halo that is not completely rejected by observations.}
 
 %Our model supports a new formation scenario for the MS: LMC-dominated tidal stripping at the onset of the recent LMC-SMC binary pair.  This MS formation scenario does not depend on a close encounter with the Milky Way and therefore is not inconsistent with more realistic longer-period L/SMC orbits.  The bound orbit in our model manifests itself in two important ways: the elongation of the LA to high galactic latitudes and the presence of filamentary substructure within the MS.
 
\end{abstract}


\section{Introduction}

The Magellanic Stream (MS) is a massive trail of HI gas which lags behind the LMC and SMC in their orbit about the Milky Way (MW).  On the leading side of the orbit is a network of HI clumps known collectively as the Leading Arm (LA), and in the region between the LMC and SMC is the Magellanic Bridge.  A number of numerical models have provided a dynamical origin for these gaseous features whereby gas is stripped away from the SMC disk by tidal gravitational forces: Murai \& Fujimoto 1980 (MF80), Gardiner \& Noguchi 1996 (GN96), Yoshizawa \& Noguchi 2003, Connors et al. 2006 (C06).  In these \emph{traditional tidal models}, the Magellanic Stream is stripped away $\sim$1.5 Gyr ago coinciding with a close encounter with the MW.

This formation scenario has recently been cast into doubt by the HST proper motion study of Kallivayalil et al. (2006a, 2006b): whereas the traditional tidal models rely on a succession of close encounters between the L/SMC and MW, the HST data seem to imply that such bound orbits are implausible (Besla et al. 2007).  This has led to a recent proposal that the MS may have formed within a first passage orbit about the MW (Besla et al 2010, hereafter B10).  The conclusion of unbound L/SMC orbits, however, strongly depends on the adopted profile of the MW halo, as demonstrated by Shattow \& Loeb (2009, hereafter SL09).  SL09 revise the MW mass profile of Besla et al. 2007 to agree with new measured values of the MW circular velocity and find that the LMC is most likely bound to the MW with a long ($\sim$6 Gyr) orbital period { (references for the increased value of $V_{\rm cir}$ are given in Table 1)}.  These conclusions must again be revised if one considers the steep MW mass profile adopted by the tidal models, i.e., an isothermal sphere.

Here we present a numerical model based upon the traditional tidal model (e.g., MF80 and GN96) but with two important differences: we incorporate proper motions for the LMC and SMC within 1$\sigma$ of the HST values, and motivated by SL09 we parametrize the MW isothermal halo with an increased circular velocity of $V_{\rm cir}=$ 250 km s$^{-1}$.  We retain an orbital scenario in which the LMC and SMC remain bound to the MW for more than 5 Gyr, and we furthermore demonstrate that structures resembling the MS, Bridge, and LA can plausibly form within the orbit.  Unlike the traditional models, however, the pericentric passage about the MW in our orbit (also occurring at $\sim$1.5 Gyr ago) does \emph{not} contribute to the tidal stripping of the SMC disk.  Instead, the tidal stripping occurs during the formation of a strong LMC-SMC binary pair $\sim$1.2 Gyr ago.  

%That is, our model suggests a new mechanism for the formation of the MS: LMC-dominated tidal forces at the onset of a short period ($\sim$1 Gyr) LMC-SMC binary pair.
%Our model is remarkably simple (e.g., neglect of self-gravity, neglect of gas physics, adoption of an isothermal sphere; see sec 4), and as such it cannot be used to study the intricate features of the Magellanic system.  We instead

Because the present model is somewhat idealized { (e.g., neglect of self-gravity, neglect of gas physics, adoption of an isothermal sphere)}, one should regard the present work as a stepping stone between the traditional models of the past and the more robust and realistic models to be developed in the future.  { We also stress that the latest proper motion measurements of the LMC, derived from the motions of 3822 stars by Vieira et al. (2010), implies that the LMC's present velocity is $\sim 340$ km s$^{-1}$ ($\pm48$ km s$^{-1}$), which is significantly lower than the earlier result ($\sim 380$ km s$^{-1}$) derived from the motions of only 810 stars by Kallivayalil et al. (2006a).  Thus, we may not need to adopt an LMC velocity as high as that of the present model ($\sim 360$ km s$^{-1}$), but we retain this value in order to demonstrate that the tidal origin of the MS can still be preserved in a bound orbit with a large LMC velocity.}

%Furthermore, we intend to outline various broad aspects of our present model which will likely survive even when the model itself is replaced.

\begin{table*}
\begin{minipage}[c]{200cm}
\caption{Observational Constraints}
\begin{tabular}{@{}lcp{6cm}}
%\tablecolumns{3}
%\tablewidth{0pc}
%\tablecaption{
%\tablehead{
\hline
Parameter & Chosen Value & References \\
\hline
\bf{Milky Way} & & \\
Circular Velocity $V_{ \rm cir }$ (km s$^{-1}$) & 250 &  Reid \& Brunthaler (2004); Uemura et al. (2000); Reid et al. (2009); Sirko et al. (2004) \\
Distance to Sun$^\itl{a}$ $R_{ \odot }$ (kpc) & 8.5 & Gillessen et al. (2009); Reid et al. (2009) \\
Velocity of Sun$^\itl{a}$  (km s$^{-1}$)& (10.0, 5.2 + $V_{ \rm cir }$, 7.2) & Dehnen \& Binney (1998) \\
\hline
\bf{LMC} & & \\
Proper Motion ($\mu_W$, $\mu_N$) (mas yr$^{-1}$) & (-2.04, 0.48) & Kallivayalil et al. (2006) \\
Line-of-Sight Velocity $v_{ \rm sys}$ (km s$^{-1}$) & 262.2 & van der Marel et al. (2002) \\
Position ($\alpha$, $\delta$) (degrees) & (81.9, -69.9) & van der Marel et al. (2002)  \\
Distance Modulus & 18.50 & Freedman et al. (2001) \\
\hline
\bf{SMC} & & \\
Proper Motion ($\mu_W$, $\mu_N$) (mas yr$^{-1}$) &  (-1.31, -1.27) & Kallivayalil et al. (2006) \\
Line-of-Sight Velocity $v_{ \rm sys}$ (km s$^{-1}$) & 146.0 & Harris \& Zaritsky (2006) \\
Position ($\alpha$, $\delta$) (degrees) & (13.2, -72.5) & Stanimirovi\'c et al. (2004) \\
Distance Modulus & 18.95 & Cioni et al. (2000) \\
\hline
$^\itl{a}$ Represented in a galactocentric frame (GSR). & & \\

\end{tabular}
\end{minipage}

\end{table*}

\section{Numerical Model}

We take our numerical framework from MF80: first, we perform a three-body orbit integration from the present day to 3 Gyr in the past, using fixed potentials and dynamical friction; and second, we evolve an exponential disk of 40,000 test particles along the previously computed SMC orbit.  Initial conditions for the orbit integration are taken from the observational constraints listed in Table 1.  Because we adopt proper motion values within 1$\sigma$ of the HST values, our present-day L/SMC velocities are much larger than those of the traditional models.  For instance, the C06 galactocentric tangential velocities are 287 km s$^{-1}$ (LMC) and 255 km s$^{-1}$ (SMC), compared to our values of 355 km s$^{-1}$ and 329 km s$^{-1}$, respectively.  
% I took out any mention of 7.5 kpc radius disk, because too cluttered!

Following the traditional tidal models, the LMC and SMC are represented by Plummer potentials (having scale radii $K_L=$ 3 kpc and $K_S=$ 2 kpc, respectively), and their masses are taken to be 2$\times10^{10}M_{\odot}$ and 3$\times10^{9}M_{\odot}$, respectively.  { These masses are consistent with the observed rotation curves derived by Kim et al. (1998) for the LMC and Stanimirovi\'c et al. (2004) for the SMC}.  Also following the traditional models, we adopt an isothermal profile for the MW halo, for which the potential is $\phi=-V_{\rm cir}^2\ln(r)$.  { This choice allows us to adopt equation 7-23 of Binney \& Tremaine (1987) for the dynamical friction experienced by the LMC and SMC as they orbit through the MW halo.}  Even though more realistic MW profiles exist (such as the NFW halo of Navarro et al. 1996), we purposely mimick the numerical framework of the traditional tidal models in order to demonstrate that the conflict with the HST proper motion data can be resolved by a simple change of parameters.

We conducted a large parameter space search over a range of values for $V_{\rm cir}$ and the proper motions $\mu_N$ $\mu_W$ of the LMC and SMC, with the goal of finding models which adequately reproduced the MS and LA.  We did indeed find a number of strong candidates, and the values of Table 1 parametrize a representative choice out of this group.  Accordingly, the results of our study are not tied down to these exact parameter choices but are valid over a flexible range.  We could not find a model which adequately reproduced the MS or LA for $V_{\rm cir}=$ 220 km s$^{-1}$, even though the traditional tidal models adopt this value.  Instead, we find that the MS and LA are more faithfully reproduced when adopting increased $V_{\rm cir}$ values, agreeing with the recent work by Ruzicka et al. (2010).  The present model incorporates a value of $V_{\rm cir}=$ 250 km s$^{-1}$, which is supported by a number of recent observational studies (see Table 1).

%Figure 1
\begin{figure}
\includegraphics[trim=0 0 30 150, width=8cm]{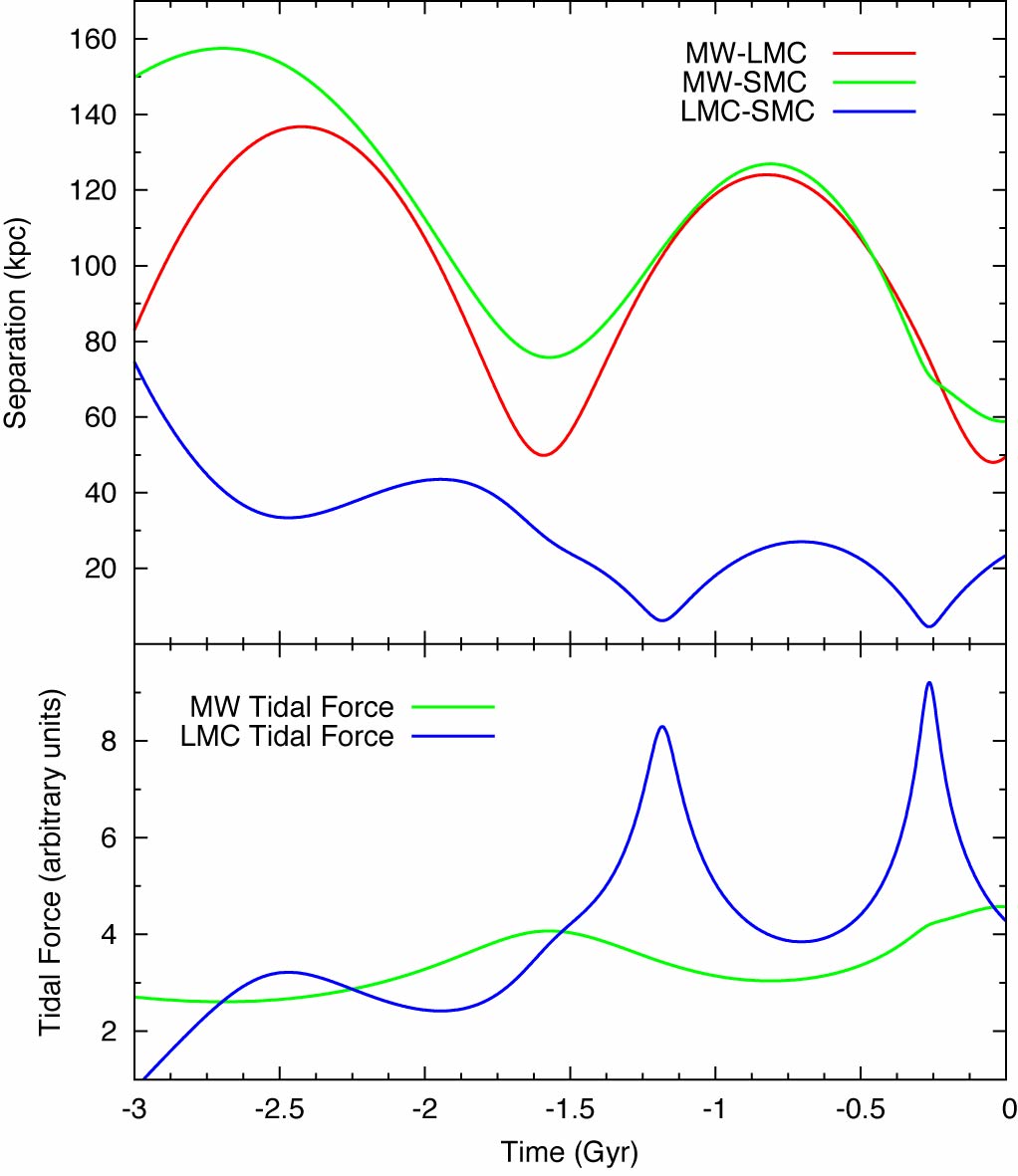}
\caption{Separations between the LMC, SMC, and MW during their 3 Gyr orbital interaction (top panel), and the tidal forces experienced by the SMC (bottom panel).}
\end{figure}

%Figure 2
\begin{figure}
\includegraphics[trim= 150 20 170 50, width=8cm]{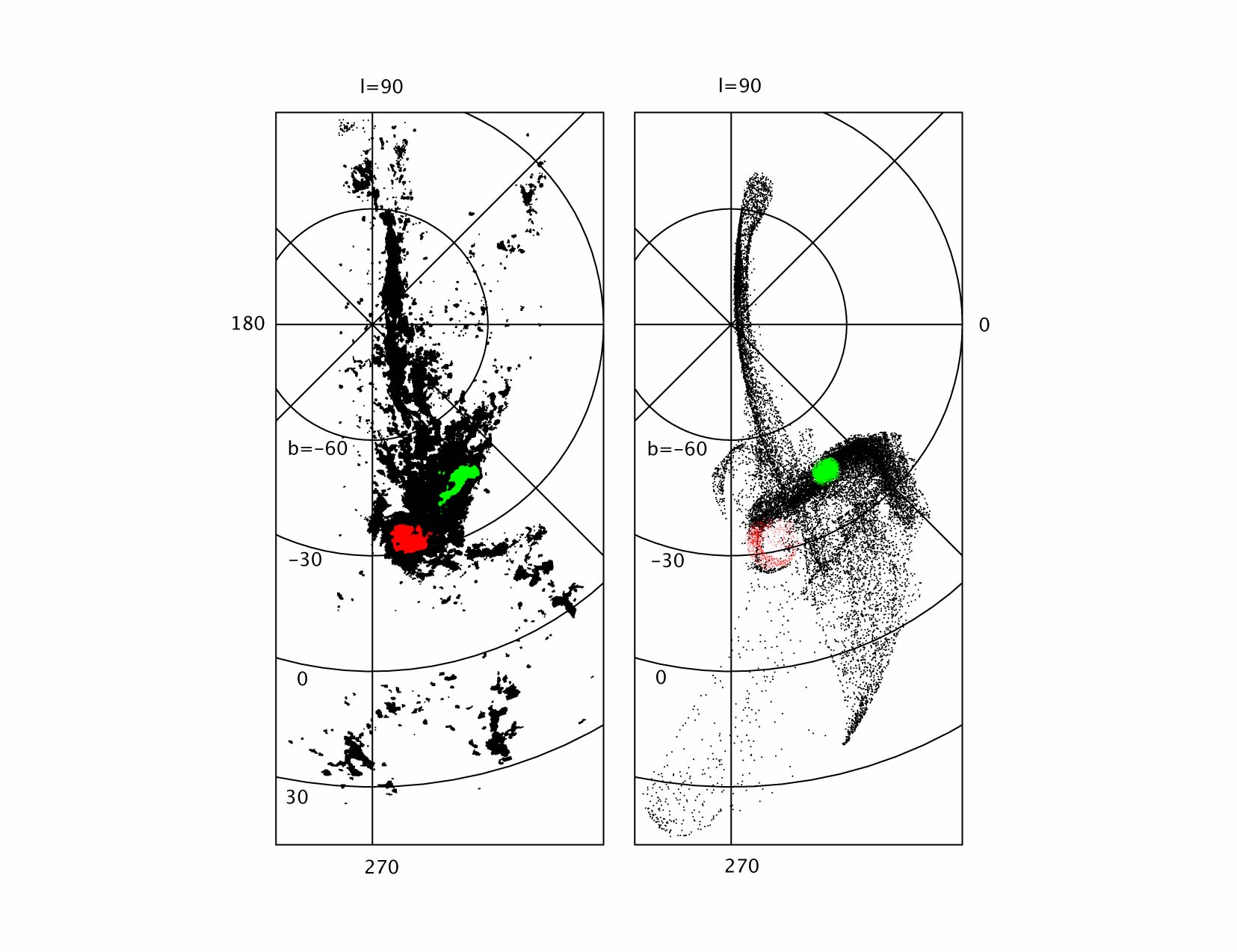}
\caption{(left panel) HI observations of Putman et al (2003) ($b<-15^{\circ}$) and Bruns et al (2005) ($b> -15^{\circ}$).  The LMC (red) and SMC (green) are shown.  (right panel) Final test particle distribution of our model as projected against the sky.  After evolving through the 3 Gyr orbital interaction, particles either remain in the SMC disk (green; i.e. within 4 kpc of the SMC), are stripped away (black), or are captured by the LMC (red; i.e., within 6 kpc of the LMC).}
\end{figure}

\section{Results}

The LMC and SMC remain tightly bound to the MW in our model (see Fig 1, top panel) with a short orbital period of $\sim$1.5 Gyr.  This bound orbit is a consequence of the steep MW mass profile implied by our adopted isothermal halo.  Within 50 kpc (i.e., the closest approach of the LMC) the enclosed mass is $\approx0.7\times10^{12}M_{\odot}$, and within 150 kpc (i.e., approximately the farthest distance of the LMC) the enclosed mass is $\approx2.2\times10^{12}M_{\odot}$.  { This MW mass profile is not ruled out by observations because constraints on the MW halo do not exist out to 160 kpc, which is the furthest extent of the LMC and SMC orbits in our model (see Figure 1).  Gnedin et al. (2010) provide observational constraints out to 80 kpc, and our MW mass is reasonable albeit high in comparison with their results to that radius.}

The binary state of the LMC and SMC is a recent phenomenon in our model, occurring only within the last $\sim$1.6 Gyr.  The LMC and SMC have executed two close passages in that time, the first 1.2 Gyr ago (coming within 6.2 kpc of each other) and again 0.25 Gyr ago (4.6 kpc).  These close encounters exposed the SMC disk to strong LMC tidal forces (see Fig 1, bottom panel) which stripped away the MS and LA from the SMC disk at the first encounter ($t=-1.2$ Gyr), and which formed the Bridge after the second encounter ($t=-0.25$ Gyr).  From Fig 1 it is clear that the LMC dominates the tidal field during the stripping epoch, in contrast to past models in which the MW also participates in the stripping process (e.g., GN96, C06).

By integrating the orbit back to 5.5 Gyr ago, we find the LMC and SMC each independently stay within 160 kpc of the MW, whereas the LMC-SMC separation exceeds 200 kpc.  This is suggestive of a scenario in which the LMC and SMC may have originally formed as independent satellites of the MW and were furthermore separated by large distances at the time of formation (Bekki et al. 2004).  In this scenario, their subsequent orbital evolution through the MW halo gradually brought them closer together until the LMC was able to capture the SMC into its orbit and form a tightly bound binary pair.  In our model, the onset of this binary pair provides the physical mechanism for the formation of the MS.

%Figure 3
\begin{figure}
\includegraphics[width=8cm]{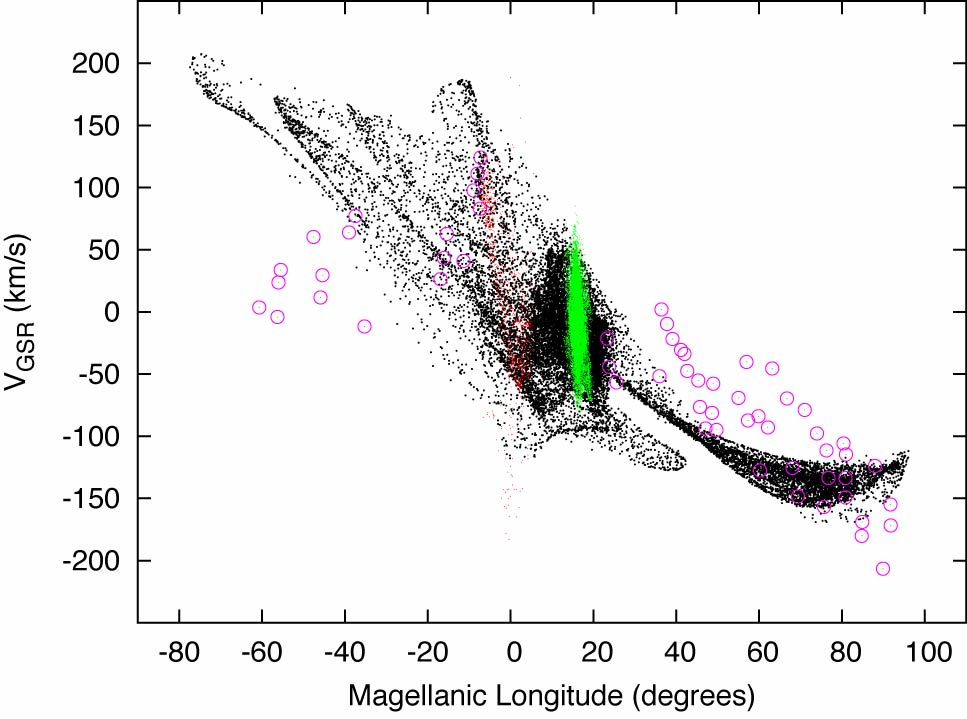}
\caption{Galactocentric radial velocities of our test particles plotted against Magellanic longitude (a coordinate which runs parallel to the MS; see Wakker 2001 for definition).  The red and green particles are those within the LMC and SMC, respectively, and the particles at right (left) constitute the MS (LA).  Circles represent observational data sampled from Bruns et al. (2005).}
\end{figure}

As seen in Figure 2, our simulated MS exhibits good positional agreement with observations, particularly the curvature at its base and the slender extent across $\sim$100$^{\circ}$ degrees in the sky.  More that 95\% of the particles in the MS originate beyond 3 kpc in the initial SMC disk, where stellar populations are thought to be low (Yoshizawa \& Noguchi 2003).  We accordingly interpret our MS as being strictly gaseous.  The approximate ratio of particles in the MS as compared to the LA is two-to-one for our model, although observations indicate a mass ratio of four-to-one (Bruns et al. 2005).  As discussed in the next section, interactions with the MW hot halo may help to resolve this discrepancy.

At positive galactic latitudes $b>0^{\circ}$, the position and extent of the LA provides a crude but promising correspondence with observations.  In fact, our model suggests a new interpretation of the LA: rather than being a single fractured structure (C06), the LA may be composed of two separate branches with distinct origins.  One branch ($l>285^{\circ}$, { colored green in Figure 5}) of the LA originates from a similar region of the disk as the MS, whereas the other branch ($l<285^{\circ}$, { red in Figure 5}) is pulled from the outermost part of the SMC disk.  { The color-coding of Figure 5 applies to Figure 6 as well, which displays the SMC disk right before it is disrupted at $t=-1.23$ Gyr.  The discrepancy in density of the two LA branches can be attributed to the fact that the low-density branch (($l<285^{\circ}$, red) arises from a much more tenuous region of the SMC disk, seen at left in Fig 6.}

During the past 0.3 Gyr, both branches of the LA are rapidly stretched to positive galactic latitudes.  It is clear that the MW drives this process: the elongation occurs in the leading direction of the L/SMC orbit about the MW, and moreover it occurs as the MW pericenter is approached.  { This process transfers considerable kinetic energy to the particles of the LA, manifested as large galactocentric velocities (Fig 3).}  Though the LA velocity profile is offset from observations by as much as 150 km s$^{-1}$, interactions with the hot halo may be able to mitigate the discrepancy, as discussed in section 4.

As first noted by Putman et al. (2003, hereafter P03), the Magellanic Stream possesses an intertwining filamentary structure across its entire length with a number of obvious spatial bifurcations.  { Such substructure within the MS is not exhibited in the on-sky projection of particles in our model (Figs 2 and 5), but it does indeed appear within the three-dimensional distribution of particles.  We have provided an animation as supplementary online material which makes this point clear (see Appendix) and which furthermore reveals that the MS is composed of two bifurcated filaments.  These filaments are not identifiable in Figs 2 and 5 simply because the on-sky projection provides an unfavorable point of view.}  As discussed below, { the MS bifurcation} in our model is in fact an imprint of { structures} which formed originally within the SMC disk.

As noted earlier, our model (hereafter, the \emph{fiducial model}) follows the evolution of the SMC disk from $t=-3$ Gyr (when it is completely undisturbed) to $t=0$ Gyr (when the MS, LA, and Bridge have fully formed).  If we instead place an undisturbed SMC disk into the orbit at $t=-1.23$ Gyr and let it evolve to the present day, we can effectively remove the influence of the interactions previous to $t=-1.23$ Gyr. In particular, the MW pericentric passage occurring 1.5 Gyr ago may be removed from the orbital dynamics in this way.  The resulting test particle distribution is shown at left in Fig 4 { and the corresponding projection for the fiducial model is shown at right in Fig 4}.

Remarkably, the global features of our fiducial model are preserved { in the $t=-1.23$ Gyr model}, including the positions, orientations, densities, and velocities of the MS, LA, and Bridge.  This nicely verifies what we stressed previously: the MW pericenter at $t=-1.5$ Gyr does \emph{not} contribute to the formation of the MS.  Noticeably different in this new model, however, is a uniform distribution of particles within the MS.  Since the interactions subsequent to $t=-1.23$ Gyr are not able to induce { bifurcation} within the MS, we must conclude that the { bifurcation} in our fiducial model was created \emph{prior} to $t=-1.23$ Gyr, when the SMC disk was still fully intact.

Before $t=-1.23$ Gyr, we see from Fig 1 that the SMC disk was influenced by comparably weak MW and LMC tidal fields.  Inspecting the state of the SMC disk { at $t=-1.23$ in Fig 6}, we find that these weak tidal fields are indeed able to disturb the disk without disrupting it, creating intricate tidal features and warps.  { In particular, the region of the SMC disk which eventually constitutes the MS (i.e. colored blue in Fig 6) exhibits non-uniform densities across a sharp ridge.}  These complex structures are subsequently imprinted into the MS when it is stripped away from the SMC disk, { providing the seed for the bifurcated filaments} in the fiducial model.  Likewise, the model which avoids this epoch of weak tidal fields accordingly imprints a uniform distribution of particles into the MS.

\section{Discussion and Conclusions}

We find that a simple change of parameters permits a new formation scenario for the Magellanic Stream within the framework of the traditional tidal models.  In particular, the MS can suitably form in a bound orbit about the Milky Way via LMC-dominated tidal stripping of the SMC disk.  The B10 model supports a similar stripping mechanism, although they adopt an unbound orbit about the MW.  Because both bound (this work) and unbound (B10) orbital scenarios appear plausible, the formation of the MS may not supply a strong condition on whether the L/SMC are bound to the Milky Way.

Nevertheless, our bound model suggests that the presence of the Milky Way has two important impacts: first, the strong MW tidal field during the most recent MW pericentric passage is able to elongate the LA to $b\approx30^{\circ}$; and second, the combined weak tidal fields of the MW and LMC from $t=-3$ Gyr to $t=-1.2$ Gyr are responsible for creating bifurcated filaments within the MS.  { Because a first passage orbit relegates the MW to a peripheral role, the B10 model failed to reproduce these two features.}  Moreover, if the observed bifurcation of the Magellanic Stream (P03) can indeed be linked to the tidal field of the MW \emph{prior} to the epoch of stripping, an obvious conflict arises with the first passage scenario.  { We unfortunately cannot assess such claims here because the present model is too simple (e.g., test particle method, neglect of gas physics) to accommodate an intricate analysis of the MS.

In our subsequent work (Diaz \& Bekki 2011) we have overcome this simplicity by extending the present model to include self-gravity and a gas-dynamical drag term.  The drag is proportional to the ram pressure encountered by the particles of the MS and LA as they orbit through the gaseous MW hot halo (Gardiner 1999).  We find that the drag force is able to reduce the total number of particles in the LA and retard their velocities, thereby mitigating the discrepancy with observations for the MS-to-LA mass ratio and the velocity profile of Fig 3, respectively.  Additionally, we find that the presence of drag enhances the on-sky bifurcation of the MS by reducing its velocity dispersion.  Though this follow-up study is only preliminary, it suggests that our model exhibits a number of promising features which deserve deeper investigation.  In particular, a fully hydrodynamical treatment may help illuminate how the LA is able to evolve from a continuous tidal feature into a discrete set of clumpy cloudlets (Bruns et al. 2005).}

%Figure 4
\begin{figure}
\includegraphics[trim=0 0 60 0, width=8cm]{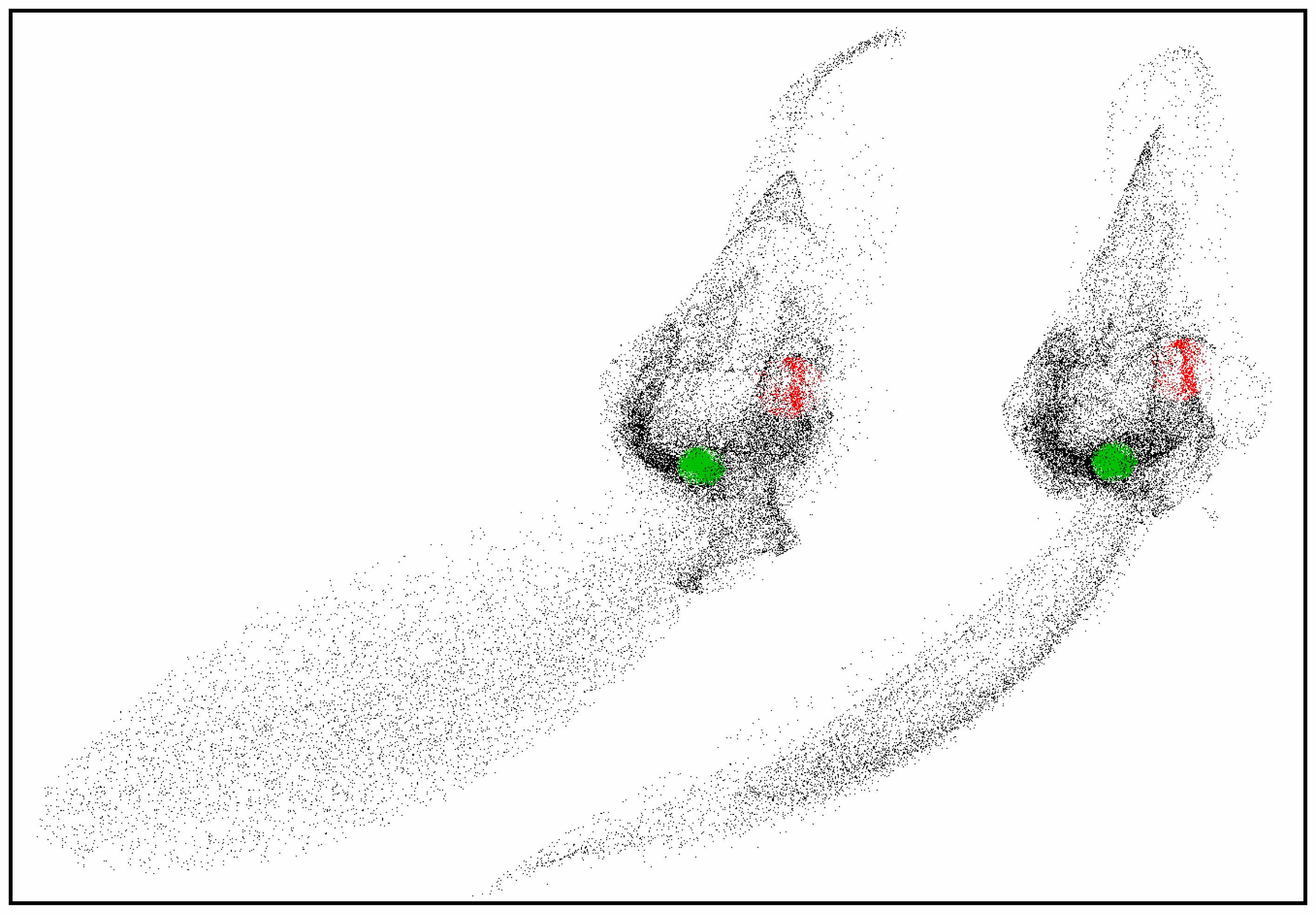}
%trim= 330 50 430 50,
\caption{Three-dimensional distribution of test particles for the model which evolves from $t=-1.23$ Gyr (left), and for the { fiducial} model which evolves from $t=-3$ Gyr (right).  The red and green particles are those within the LMC and SMC, respectively, and the lower and upper features are the MS and LA, respectively.  { Structure} within the MS is evident for the model at right and noticeably absent in the model at left.  The bounding box has a vertical dimension of 170 kpc.}
\end{figure}

%Figure 5
\begin{figure}
\begin{centering}
\includegraphics[width=4cm]{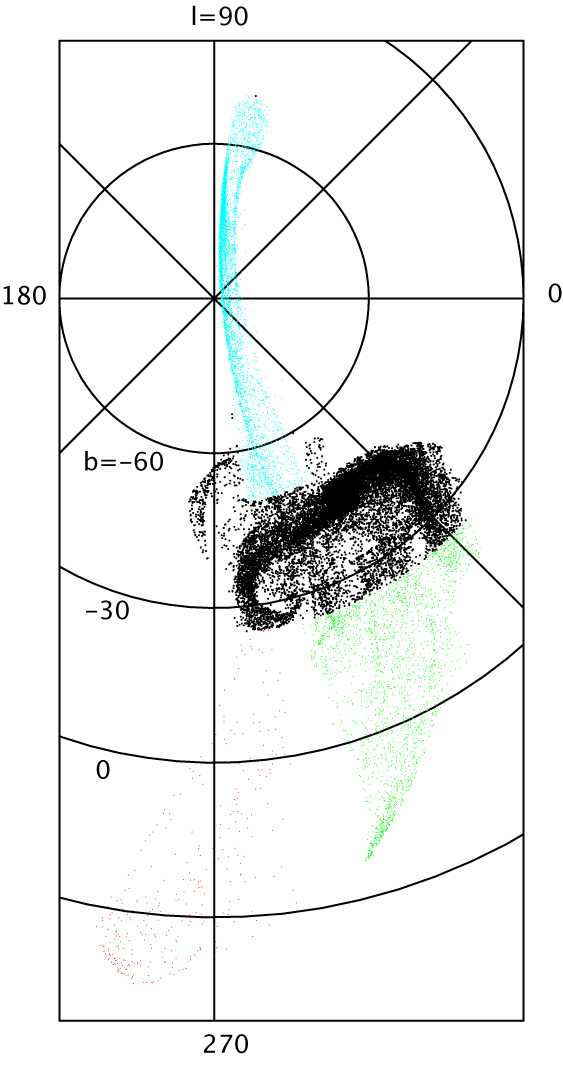}
\caption{Color-coding which highlights the individual components of the final test particle distribution of the fiducial model.  The particles constituting the MS (blue) and the two branches of the LA (green $l>285^{\circ}$; red $l<285^{\circ}$) are shown separately.  All other particles are colored black.}
\end{centering}
\end{figure}

%Figure 6
\begin{figure}
\begin{centering}
\includegraphics[width=8cm]{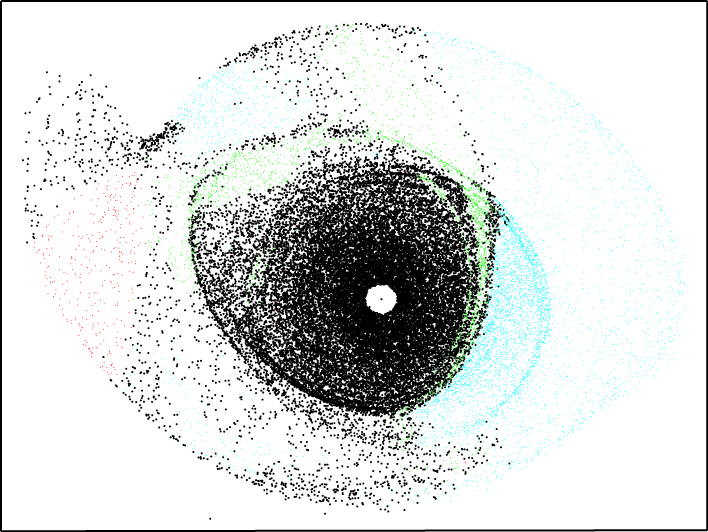}
\caption{Color-coding of the undisrupted SMC disk at $t=-1.23$ Gyr, indicating the particles which will eventually constitute the MS (blue) and the two branches of the LA (green $l>285^{\circ}$; red $l<285^{\circ}$).  All other particles are colored black.  The bounding box has a vertical dimension of 16 kpc.}
\end{centering}
\end{figure}

Although we chose an isothermal MW halo to purposefully imitate the traditional tidal models, this choice threatens the credibility of our model.  The isothermal halo produces a steep mass profile which artificially ensures that the rotation curve of the MW remains flat.  These conditions are not very realistic, especially at the large radii (50 kpc to 150 kpc) probed by the LMC orbit about the MW.  A more reasonable choice for the MW halo is, for instance, the NFW halo adopted by SL09.  As a consequence, the SL09 orbits are more plausible: they are bound but with considerably longer periods about the MW ($\sim$6 Gyr versus $\sim$1.5 Gyr in our model).

As discussed in B10, the traditional scenario for the formation of the MS (e.g., GN96) is incompatible with these long-period orbits.  However, we have proposed a stripping mechanism which is independent of the MW pericenter and in principle does not conflict with the long-period orbits of SL09, i.e., strong tidal forces associated with the formation of a recent LMC-SMC binary pair.  In this sense, our proposed MS formation scenario will likely survive even when the present model is replaced with more realistic MW halos and more plausible orbits.
%As long as the LMC can capture the SMC into its orbit at recent times, the orbit about the MW may not matter to the MS formation.

While our model and the B10 model both support an LMC-dominated stripping mechanism, there is a clear difference in the supplementary role played by the MW.  In our bound model, the MW mediates the binary action of the LMC-SMC pair, guiding them into a recently formed short-period orbit.  In the unbound B10 model, the binary action of the LMC and SMC is predetermined as a two-body system in isolation of the MW.  The particular LMC-SMC orbit of the B10 model is furthermore curious in that the SMC is presently \emph{approaching} the LMC.  This is in apparent contradiction with the observed LMC-SMC relative velocity (Kallivayalil et al. 2006b).
%It is also unclear whether the B10 model is consistent with the present-day positions and velocities of the LMC and SMC.  For example, the B10 model does not incorporate the HST proper motion of the SMC.

Our orbital model - in particular the formation of the recent LMC-SMC binary pair - may provide insights into a number of outstanding research questions.  Harris \& Zaritsky (2009) have determined that the star formation histories of the LMC and SMC are coupled, with two correlated epochs of star formation occurring $\sim$2 Gyr ago and $\sim$0.5 Gyr ago.  { Although the timing of the two close LMC-SMC passages in our model are not exactly consistent with the timing of these burst epochs, the present study may well imply that such starbursts could be associated with two strong tidal interactions between the LMC and SMC during their recent dynamical coupling.  Our orbital model may also naturally explain why star formation fell into a quiescent epoch from 12 Gyr to 5 Gyr ago (Harris \& Zaritsky 2009) because the LMC and SMC were separated by large distances and were unable to induce star formation in their respective disks.}

The formation of a recent LMC-SMC binary pair has also been proposed by Bekki et al. (2004) to explain the unique distribution of ages among globular clusters of the LMC.  { Most of the LMC globular clusters were formed either at ancient times ($\sim$13 Gyr ago) or at recent times ($< 3$ Gyr ago) (Da Costa 1991).  This puzzling $\sim$10 Gyr age gap was explained by Bekki et al. (2004) by invoking a recent epoch of globular cluster formation induced by strong interactions between the LMC and SMC.}  Though the orbital model of Bekki et al. (2004) is inconsistent with HST proper motion data, our present model indicates that a similar orbital scenario is still possible.

%The two close LMC-SMC passages in our model may provide the physical justification for such a coupling, as one can imagine star formation igniting when the mutual LMC-SMC tidal interaction causes a simultaneous compression of their gaseous disks.  
%Furthermore, our model explains why there are only \emph{two} such correlated epochs of star formation, because the LMC and SMC were not bound to one another previous to these encounters.
 
Drawing upon the framework of the traditional models, we have developed a well-constrained tidal model which reconciles { the latest proper motion data (e.g. Kallivayalil et al. 2006a, 2006b; Vieira et al. 2010)} with the formation of the Magellanic Stream in a bound orbit.  Our model furthermore suggests that the LMC and SMC may have been separate entities when they formed, and that many of their observed properties, including the existence of the MS and LA, can be attributed to their recent activity as a binary pair.

{ Even though the LMC velocity adopted by the traditional tidal models (e.g., $\sim 300$ km s$^{-1}$ for GN96 and C06) are inconsistent with the HST proper motion data, they are still valid to within 1-$\sigma$ error of the LMC velocity $\sim 340 \pm 48$ km s$^{-1}$ implied by Vieira et al. (2010).  Taken together with our present model, these results indicate that bound orbits for the LMC and SMC are still viable in self-consistently explaining the properties of the MS and LA.  Furthermore, we have shown that the MW circular velocity adopted by the traditional models $V_{\rm cir}=$ 220 km s$^{-1}$ must be increased in order to retain a bound orbital history.  Ruzicka et al. (2010) come to a similar conclusion, thus providing another piece of evidence that the higher velocities of the LMC and SMC may still be consistent with a bound MS formation scenario.  Lastly, these tidal models rely on the key assumption that the MS originated in the SMC disk, and the recent abundance study of Fox et al. (2010) provides a strong observational justification for this assumption: the low metallicity at the tip of the MS is consistent with metallicities observed in the SMC disk but \emph{not} the LMC.}
%{\bf We suggest that if the present-day velocity of the LMC is large ($>$300 km s$^{-1}$), then the MW circular velocity must be larger than 220 km s$^{-1}$ in order to self-consistently reproduce the observed properties of the MS and LA.}

%\acknowledgments
JD is supported by a Sir Keith Murdoch Fellowship and a SIRF scholarship.  We are grateful to the anonymous referee for constructive and useful comments.  We thank Lister Staveley-Smith for thoughtful discussions and access to data used in Figs 2 and 3.

\appendix
\section{MS Bifurcation}

As supplementary online material we have provided an animation file a1.mp4 which gives the three-dimensional locations of particles in the MS and LA for our fiducial model (at right) and for the model which evolves from $t=-1.23$ Gyr (at left).  The animation is able to depict three-dimensional positions by rotating the test particle distributions through a full 360$^{\circ}$.  Figure 4 is adapted from a single frame of the animation.

The bifurcation of the MS in our fiducial model is readily identifiable in the animation.  At various angles of rotation, the MS clearly splits into two separate filaments.  This bifurcation unfortunately does not map directly to the on-sky projection and thus is not visible in Figures 2, 4, or 5.

The model which evolves from $t=-1.23$ Gyr exhibits the same spatial and kinematical features as our fiducial model, and the animation shows this clearly for the spatial distributions of the MS, LA, and Bridge.  Remarkably, the animation shows that only difference between the two models is the substructure of the MS.  Whereas the MS of the fiducial model is clearly bifurcated, the MS of the model which evolves from $t=-1.23$ Gyr exhibits a uniform distribution of particles.  We accordingly interpret the MS bifurcation to be the result of weak tidal interactions between $t=-3$ Gyr and $t=-1.23$ Gyr when the SMC disk was still intact.


\begin{thebibliography}{}

\bibitem[]{} Bekki K., et al. 2004, ApJ, 610, L93

\bibitem[]{} Besla G., Kallivayalil N., Hernquist L., Robertson B., Cox T. J., van der Marel R. P., \& Alcock C., 2007, ApJ, 668, 949

\bibitem[]{} Besla G., Kallivayalil N., Hernquist L., van der Marel R. P., Cox T. J., \& Keres D., 2010, ApJ, 721, L97 (B10)

\bibitem[]{} Binney J., \& Tremaine, S., 1987, Galactic Dynamics. Princeton Univ. Press, Princeton, NJ

\bibitem[]{} Bruns C., et al., 2005, A\&A, 432, 45

\bibitem[]{} Cioni M.-R. L., van der Marel R. P., Loup C., \& Habing H. J., 2000, A\&A, 359, 601

\bibitem[]{} Connors T. W., Kawata D., \& Gibson B. K., 2006, MNRAS, 371, 108 (C06)

\bibitem[]{} Da Costa G. S. 1991, in IAU Symp. 148, The Magellanic Clouds, ed. R. Haynes \& D. Milne (Dordrecht: Kluwer), 183

\bibitem[]{} Diaz J., \& Bekki K., 2011, submitted to PASA

\bibitem[]{} Dehnen W., \& Binney J. J., 1998, MNRAS, 298, 387

\bibitem[]{} Fox, A., et al. 2010, ApJ, 718, 1046

\bibitem[]{} Freedman W. L., et al. 2001, ApJ, 553, 47

\bibitem[]{} Gardiner L. T., \& Noguchi M., 1996, MNRAS, 278, 191 (GN96)

\bibitem[]{} Gardiner L. T., 1999, in ASP Conf. Ser. 166, Stromlo Workshop on High-Velocity Clouds, ed. B. K. Gibson \& M. E. Putman (San Francisco: ASP), 292

\bibitem[]{} Gillessen S., et al. 2009, ApJ, 692, 1075

\bibitem[]{} Gnedin, O. et al., 2010, ApJ, 720, L108

\bibitem[]{} Harris J., \& Zaritsky D., 2006, AJ, 131, 2514

\bibitem[]{} Harris J., \& Zaritsky D., 2009, AJ, 138, 1243

\bibitem[]{} Kallivayalil N., van der Marel R. P., Alcock C., Axelrod T., Cook K. H., Drake A. J., \& Geha M., 2006, ApJ, 638, 772 (2006a)

\bibitem[]{} Kallivayalil N., van der Marel R. P., \& Alcock C., 2006, ApJ, 652, 1213 (2006b)

\bibitem[]{} Kim, S. et al. 1998, ApJ, 503, 674

%\bibitem[]{} Klypin A., Zhao H., \& Somerville R. S., 2002, ApJ, 573, 597

%% not referenced in the body of the paper
%\bibitem[]{} Mastropietro, C., Moore, B., Mayer, L., Wadsley, J., \& Stadel, J., 2005, MNRAS, 363, 509

\bibitem[]{} Murai T., \& Fujimoto M., 1980, PASJ, 32, 581 (MF80)

\bibitem[]{} Navarro J. F., Frenk C. S., \& White S. D. M., 1996, ApJ, 490, 493

%% not referenced in the body of the paper
%\bibitem[]{} Piatek, S., Pryor, C., \& Olszewski, E. W., 2008, ApJ, 135, 1024

\bibitem[]{} Putman M. E., Staveley-Smith L., Freeman K. C., Gibson B. K., \& Barnes D. G., 2003, ApJ, 586, 170 (P03)

\bibitem[]{} Reid M. J., \& Brunthaler A., 2004, ApJ, 616, 872

\bibitem[]{} Reid M. J., et al. 2009, ApJ, 700, 137

\bibitem[]{} Ruzicka A., Theis C., \& Palous J., 2010, arXiv:1010.0942, ApJ in press

\bibitem[]{} Shattow G., \& Loeb A., 2009, MNRAS, 392, L21 (SL09)

\bibitem[]{} Sirko E., et al. 2004, AJ, 127, 914S

\bibitem[]{} Stanimirovi\'c S., Staveley-Smith L., \& Jones P. A., 2004, ApJ, 604, 176

\bibitem[]{} Uemura M., Ohashi H., Hayakawa T., Ishida E., Kato T., \& Hirata R., 2000, PASJ, 52, 143

\bibitem[]{} van der Marel R. P., Alves D. R., Hardy E. \& Suntzeff N. B., 2002, AJ, 124, 2639

\bibitem[]{} Vieira et al. 2010, arXiv1009.4218, AJ in press

\bibitem[]{} Wakker B. P., 2001, ApJS, 136, 463

\bibitem[]{} Yoshizawa A. M., \& Noguchi M., 2003, MNRAS, 339, 1135

\end{thebibliography}
\end{document}